\documentclass[11pt]{article}
\usepackage{blois}
\usepackage{amssymb,amsmath,amsthm,bm}

\def\Journal#1#2#3#4{{#1} {\bf #2}, #3 (#4)}

\def\NIMA{{\em Nucl. Instrum. Methods} A}

\def\PLB{{\em Phys. Lett.}  B}
\def\PRL{\em Phys. Rev. Lett.}
\def\PRD{{\em Phys. Rev.} D}

\def\be{\begin{equation}}
\def\ee{\end{equation}}
\def\bea{\begin{eqnarray}}
\def\eea{\end{eqnarray}}

\newcommand{\Photo}{}

\begin{document}
\vspace*{4cm}
\title{NEW CONSTRAINTS ON NEUTRINO MILLICHARGE FROM TERRESTRIAL EXPERIMENTS AND ASTROPHYSICS}

\author{A. STUDENIKIN$^{a,b}$, I. TOKAREV$^a$}

\address{
$^a$Department of Theoretical Physics, Moscow State University, 119992 Moscow, Russia\\
$^b$Joint Institute for Nuclear Research, Dubna 141980, Moscow Region, Russia
}

\maketitle\abstracts{Phenomenology of possible neutrino millicharge in
terrestrial and astrophysical settings is discussed.
Two new limits on the millicharge from terrestrial
experiments and astrophysics are reported.}

\section{Introduction}

Neutrino electromagnetic properties are among the most intriguing and exciting problems in modern particle physics. Within the Standard Model in the limit of massless neutrinos the particle electromagnetic properties vanish. However, in different extensions of the Standard Model a massive neutrino has nontrivial electromagnetic properties (for a review of the neutrino electromagnetic properties see \cite{Giunti:2008ve,Broggini:2012df}). That is why it is often claimed that neutrino electromagnetic properties open ``a window to the new physics" \cite{Studenikin:2008bd}.

A neutrino magnetic moment, as expected in the easiest generalization of the Standard Model, is very small and proportional to the neutrino mass $m$, $\mu_\nu \approx 3 \times 10^{-19}\mu_B(m/1~{\rm eV})$ \cite{Fujikawa:1980yx} where $\mu_B=e_0/(2m_e)$ and  $e_0$ and $m_e$ are the absolute value of the
electron charge and the electron mass. Much greater values are predicted in other various Standard Model generalizations (for details see \cite{Giunti:2008ve,Broggini:2012df}). It is also usually believed that a neutrino has no electric charge. This can be attributed to gauge invariance and anomaly cancelation constraints imposed in the Standard Model. However, if a neutrino has a mass, the statement that the neutrino electric charge is zero is not so evident as it may seem to be. In theoretical models with the absence of hypercharge quantization the electric charge gets ``dequantized'' \cite{Foot:1990uf,Foot:1992ui} as well and as a result neutrinos may become electrically millicharged particles. Note that only a Dirac neutrino may have the electric charge.

In the terrestrial experiments the most severe experimental constraints on the magnetic moment $\mu_{\nu}\leq2.9\times10^{-11}\mu_B$ \cite{Beda:2012zz} was obtained from the measurements of the reactor antineutrino scattering on electrons performed by the GEMMA collaboration. The upper bound on the absolute value of the neutrino electric charge $q_0\leq10^{-21}e_0$ \cite{Marinelli:1983nd,Baumann:1988ue} was derived from the neutrality of matter and neutron itself under the assumption of the electric charge conservation.  The best astrophysical constraint on the magnetic moment, $\mu_{\nu}\leq 3.2\times10^{-11}\mu_B$ \cite{Raffelt}, was obtained from the consideration of the red gaint stars cooling .  In astrophysics the best model-independent limit on the neutrino electric charge was derived from the absence of an anomalous dispersion of the SN 1987A neutrino signal which leds to $q_0\leq3\times10^{-17} e_0$ \cite{Raffelt,Barbiellini:1987zz}.

\section{Neutrino millicharge in GEMMA experiment}

Now we consider the direct constraints on the neutrino millicharge obtained using data on the neutrino electromagnetic cross section in the GEMMA experiment.
The prescription to obtain the bound on the neutrino millicharge from
the experimental data on the ${\bar{\nu}}-e$ cross section is as follows \cite{Studenikin:2013my}. One first compares the magnetic moment cross section $\left(\frac{d\sigma}{dT}\right)_{\mu^{a}_{\nu}}$ with the Standard Model weak contribution to the cross section $\left(\frac{d\sigma}{dT}\right)_{weak}$. From the fact that the   experimental data on the cross section, for the presently achieved electron recoil energy threshold $T$, shows no deviation from the predictions of the Standard Model a limit on the neutrino magnetic moment is obtained. Then one should compare the magnetic moment contribution to the cross section, $\left(\frac{d\sigma}{dT}\right)_{\mu^{a}_{\nu}}$, and the contribution due to the neutrino millicharge, $\left(\frac{d\sigma}{dT}\right)_{q_{\nu}}$, and account that the later is also not visible at the present experiment. In order not to contradict to the experimental data the cross section $\left(\frac{d\sigma}{dT}\right)_{q_{\nu}}$ should not accede the cross section $\left(\frac{d\sigma}{dT}\right)_{\mu^{a}_{\nu}}$ anyway. Thus, the obtained upper limit on the neutrino millicharge depends on the achieved upper limit on the neutrino (anomalous) magnetic moment and the electron recoil energy threshold of the ${\bar{\nu}}-e$ experiment.

Consider the latest results \cite{Beda:2012zz} of the GEMMA collaboration on the neutrino magnetic moment. Within the presently reached electron recoil energy threshold of
$T \sim 2.8 \ keV$
the neutrino magnetic moment is bounded from above by the value
$\mu_{\nu}^{a} < 2.9 \times 10^{-11} \mu_{B}$.
In order to get from these data the limit on the neutrino millicharge we compare the mentioned above two cross sections, $\left(\frac{d\sigma}{dT}\right)_{\mu^{a}_{\nu}}$ and $\left(\frac{d\sigma}{dT}\right)_{q_{\nu}}$. The expression for the neutrino magnetic moment cross section can be found in \cite{Vogel:1989iv}, for our present needs only the term proportional to $\frac{1}{T}$ matters,
\begin{equation}\label{sigma_mu_1_T}
\left(\frac{d\sigma}{dT}\right)_{\mu^{a}_{\nu}} \approx
\pi\alpha^{2}\frac{1}{m_{e}^{2}T}
\left(\frac{\mu^{a}_{\nu}}{\mu_{B}}\right)^{2},
\end{equation}
here $\alpha $ is the fine structure constant. For the corresponding neutrino millicharge-to-charge cross section we obtain (see also  \cite{Berestetskii:1979aa})
\begin{equation}\label{sigma_q_e}
\left(\frac{d\sigma}{dT}\right)_{q_{\nu}}\approx 2\pi\alpha
\frac{1}{m_{e}T^2}q_0^2.
\end{equation}
For the ratio $R$ of the mentioned above cross sections (\ref{sigma_q_e}) and (\ref{sigma_mu_1_T}) we have
\begin{equation}\label{R}
R=\frac{\left(\frac{d\sigma}{dT}\right)_{q_{\nu}}}
{\left(\frac{d\sigma}{dT}\right)_{\mu^{a}_{\nu}}}=
\frac{2 m_e}{T}\frac{\left(\frac{{q}_0}{e_0}\right)^{2}}
{\left(\frac{\mu^{a}_{\nu}}{\mu_{B}}\right)^{2}}.
\end{equation}
In case there are no observable deviations from the weak contribution to the neutrino scattering cross section it is possible to get the upper bound for the neutrino millicharge demanding that possible effect due to $q_{\nu}$ does not exceed one due to the neutrino (anomalous) magnetic moment. This implies that $R<1$ and from (\ref{R}) we get
\begin{equation}\label{q_limit}
q_0^{2}<\frac{T}{2m_e}\left(\frac{\mu^{a}_{\nu}}{\mu_{B}}\right)^{2}e_0.
\end{equation}
Thus, from the present GEMMA experiment data and the upper limit on the neutrino millicharge is set on the level
\begin{equation}\label{q_2012}
q_0 < 1.5 \times 10^{-12} e_0.
\end{equation}
It is interesting to estimate the range of the neutrino millicharge that can be probed in a few years with the GEMMA-II experiment that is now in preparation and is expected to get data in 2015. It is planed (for details see in \cite{Beda:2012zz}) that the effective threshold will be reduced to $T=1.5 \ keV$ and the sensitivity to the neutrino anomalous magnetic moment will be at the level $1\times 10^{-11} \mu_{B}$. Then in case no indications for effects of new physics were observed from (\ref{q_limit}) we predict that the upper limit on the neutrino millicharge will be
\begin{equation}\label{q_bound_GEMMA_2}
q_0 < 3.7 \times 10^{-13} e_0 .
\end{equation}

Now it is also discussed the perspectives of the GEMMA-III experiment aimed to reach the threshold $T= 400 \ eV$ and the sensitivity to $\mu_{\nu}$ at the level $9\times 10^{-12} \mu_{B}$ approximately to the year 2017. Then  if again there were no deviations from the Standard Model cross section observed the upper limit to the neutrino millicharge will be
\begin{equation}\label{q_bound_GEMMA_3}
q_0 < 2 \times 10^{-13} e_0 .
\end{equation}
The bound on the neutrino millicharge
(\ref{q_bound_GEMMA_3}) will be reached irrespectively of whether any deviation from the
Standard Model in the cross section ${\bar {\nu}}-e$ were observed or not in the future experiments.
\section{Millicharged neutrinos in astrophysics}

There are a lot of potential sources of strong neutrino beams in astrophysics (core collapses of Supernovae, pulsars, gamma-ray bursts, active galactic nuclei, micro-quasars, etc.). A peculiar feature of the sources is a presence of high density matter and strong magnetic fields which effect on the neutrino propagation due to weak and electromagnetic interactions. Such conditions are most suitable for possible manifestations of the nontrivial electromagnetic properties of a neutrino. Here we consider the millicharged neutrino moving in matter and in
the presence of a constant magnetic field within the so called ``method of exact solutions" that implies the use of exact solutions of the modified Dirac equations \cite{StudTern,Stud2008} for the neutrino wave function.
In the considered bellow case the modified Dirac equation has the following form (see also
\cite{Balantsev:2010zw} \cite{Balantsev:2012ep} \cite{Balantsev:2013aya}),
\begin{equation}
\label{quantum_equation}
\left(\gamma_{\mu}P^{\mu}-\frac12\gamma_{\mu}(1+\gamma_5)f^{\mu}-m\right)\Psi(x)=0,
\end{equation}
where $P^{\mu}=p^{\mu}+q_0A^{\mu}$ is a neutrino kinetic momentum and $f^{\mu}$ the neutrino potential in the background neutron matter. We also account for the possible effect of matter rotation and consider a particular
case the vectors of matter rotation frequency $\bm\omega$ and of the magnetic field $\bm B$
are aligned along the third coordinate axis $\bm e_z$ thus $f^{\mu}=-Gn_n(1, -y\omega, x\omega, 0)$ and $A^{\mu}=(0,-\frac{yB}{2},\frac{xB}{2},0)$ correspondingly ($n_n$ is the neutron density, $G=\frac{G_F}{\sqrt{2}}$, $G_F$ is the Fermi constant, $\omega$ is the matter rotation frequency). The solution of Eq.~(\ref{quantum_equation}) yields discrete energy spectrum of a millicharged neutrino in the dense magnetized matter
\begin{equation}
\label{energy_spectrum}
p_0=\sqrt{p_3^2+2N(2Gn_n\omega+q_0B)+m^2}-Gn_n,
\end{equation}
where $N=0,1,2..$ is the number of a modified Landau level. Within a quasi-classical interpretation
the neutrino energy states~(\ref{energy_spectrum}) originate due to the action of an effective force \cite{Stud2008} that is produced by both weak and electromagnetic interactions of a millicharged neutrino with the dense magnetized rotating matter and has the form
\begin{equation}
\label{force}
\bm F=-(2Gn_n\omega+q_0B) \left[\bm\beta\times\bm{e}_z\right],
\end{equation}
where $\bm\beta$ is a neutrino velocity. The effective force~(\ref{force}) exemplifies the interconnection of these two types of fundamental interactions and is not vanished even in the case of a zero neutrino electric charge. This force seems to be very small for any reasonable choice of the background
parameters but it can play an important role for neutrinos in certain astrophysical settings \cite{Studenikin:2012vi}. In particular, during a supernova core collapse escaping neutrinos can be deflected on an angle
\begin{equation}
\Delta\phi \simeq \frac{R_S}{R}\sin\theta, \quad R=\sqrt{\frac{2N}{(2Gn_n\omega+q_0B)}},
\end{equation}
where $R_S$ is the radius of the star, $R$ is the radius of the neutrino trajectory and $\theta$ is an azimuthal angle of neutrino propagation. We predict that initially coincided light and neutrino beams will be spatial separated after passing through a dense rotating magnetized matter. In this connection in terrestrial experiments joint observations of initially coincided light and neutrino signals from an astrophysical transient source should not occur due to their spatial separation $\Delta L\simeq\Delta\phi L$ ($L$ is distance to the source). This new effect can explain the recent experimental results of the ANTARES experiment \cite{antares}.

On the other hand the feedback of the effective force~(\ref{force}) from the escaping neutrinos to the star matter should effect the star evolution. In particular, the torque produced by the escaping neutrinos shifts the star angular frequency
\begin{equation}
\label{delta_omega}
|\triangle\omega|=\frac{5N_{\nu}}{6M_S}(2Gn_n\omega+q_0B),
\end{equation}
where $\triangle\omega=\omega-\omega_0$ ($\omega_0$ is an initial star rotation frequency), $M_S$ is the star mass and $N_{\nu}$ is the number of the escaping neutrinos. We have termed the phenomenon as ``Neutrino Star Turning'' ($\nu ST$) mechanism. Note that depending on the neutrino millicharge sign the star rotation due to the $\nu ST$ mechanism can either spin up ($\triangle\omega>0$ for $q_{\nu}>0$) or spin down ($\triangle\omega<0$ for $q_{\nu}<0$).

The value of the relative rotation frequency shift~(\ref{delta_omega}) recalls sporadical sudden increase of a pulsar rotation frequency (a pulsar glitch \cite{glitch}). The obtained results are very important for astrophysics in light of the recent observed ``anti-glitch'' event \cite{antiglitch} that is sudden decrease of a pulsar rotation frequency. The $\nu ST$ mechanism can be used to explain both glitches and ``anti-glitches'' as well.

It is of particular interest to estimate the impact of the $\nu ST$ mechanism on a pulsar rotation rate during its formation in a supernova explosion when a great number of neutrinos $N_{\nu} \sim 10^{58}$ is produced~\cite{SN1987A}.
In case of a zero neutrino millicharge the $\nu ST$ mechanism is produced only due to weak interactions and yields
\begin{equation}
\label{delta_omega_weak}
\frac{|\triangle\omega|}{\omega_0}=4\times10^{-66}N_{\nu}
\left(\frac{1.4M_{\odot}}{M_{S}}\right)
\left(\frac{\rho_n}{10^{14}\textrm{g}/\textrm{cm}^3}\right),
\end{equation}
where $\rho_n$ is the number density of the neutron matter and we have considered the pulsar with mass $M_S = 1.4  M_{\odot}$ ($M_{\odot}$ is the Solar mass). In case of a nonzero neutrino millicharge electromagnetic interactions provide the dominant contribution to the $\nu ST$ mechanism ($q_0B\gg2Gn_n\omega$) and one can neglect weak interactions of neutrinos with the background matter. From Eq.~(\ref{delta_omega}) we get
\begin{equation}
\label{delta_omega_nonzero_charge}
\frac{|\triangle\omega|}{\omega_0}=7.6\times
10^{18}
\left(\frac{q_0}{e_0}\right)
\left(\frac{P_0}{10\text{ s}}\right)
\left(\frac{N_{\nu}}{10^{58}}\right)
\left(\frac{1.4  M_{\odot}}{M_{S}}\right)
\left(\frac{B}{10^{14} G}\right),
\end{equation}
where $P_0$ is a pulsar initial spin period. The current pulsar timing observations \cite{pulsar_timing} show that the present-day rotation periods are up to $10$ s. The rotation during the life of a pulsar spins down due to several various mechanisms and dominantly due to a magnetic dipole braking. However, all of the estimations of feasible initial pulsars spin periods give the values that are very close to the present observed periods. Thus, the estimation~(\ref{delta_omega_nonzero_charge}) is given for $P_0=10$ s. The possible existence of a nonzero negative neutrino millicharge should not significantly change the rotation of a born pulsar. From the straightforward demand $|\triangle\omega| < \omega_0$ and Eq.~(\ref{delta_omega_nonzero_charge}) we obtain the upper limit on the neutrino millicharge
\begin{equation}
\label{bound_q_nu}
q_0<1.3\times10^{-19} e_0.
\end{equation}
That is, in fact, one of the most severe astrophysical limits on the neutrino millicharge \cite{Raffelt}.

\section{Conclusions}

We consider the effects originated by millicharged neutrinos in terrestrial and astrophysical observations and derive two new limits on the neutrino electric millicharge. The first limit $q_0 < 1.5 \times 10^{-12} e_0$ is obtained from the consideration of the recently reported GEMMA collaboration results on the neutrino electromagnetic cross section. We also predict that in a few years this limit will be improved to the level $q_0 < 2 \times 10^{-13} e_0$ after the expected progress of the GEMMA collaboration  measurements  of the antineutrino scattering on electrons were achieved. The second one is based on the theoretical investigation of the millicharged neutrinos in dense magnetized matter (these background conditions are peculiar for astrophysics). In particular, we considered supernova neutrinos propagating inside the star and predicted two new phenomena: a new effect of the neutrino deviation by the rotating matter and a new mechanism of the star rotation frequency shift (the $\nu ST$ mechanism). The $\nu ST$ mechanism has significant phenomenological consequences for supernova neutrinos and the star evolution itself and yields a new astrophysical limit on the neutrino millicharge $q_0 < 1.3\times10^{-19} e_0$.

\section*{Acknowledgments}

Authors are thankful to Jean Tran Thanh Van for the kind invitation to participate at the 25rd Rencontres de Blois and to all of the organizers for their hospitality in Blois. This study has been partially supported by the Russian Foundation for Basic Research (grants No. 11-02-01509, 14-02-00914 a and 14-02-31816 mol\_a) and the Ministry of Education and Science of Russia (state contracts No. 8423 and 8415).


\section*{References}

\begin{thebibliography}{99}

\bibitem{Giunti:2008ve}C. Giunti and A. Studenikin, {\it Phys. Atom. Nucl.} \textbf{72}, 2089 (2009).

\bibitem{Broggini:2012df}C. Broggini {\it et al}, {\it Adv. High Energy Phys.} \textbf{2012}, 459526 (2012).

\bibitem{Studenikin:2008bd}A. Studenikin, {\it Nucl. Phys. Proc. Suppl.} \textbf{188}, 220 (2009).

\bibitem{Fujikawa:1980yx}K. Fujikawa and R. Shrock \Journal{\PRL}{45}{963}{1980}.

\bibitem{Foot:1990uf}R. Foot {\it et al}, {\it Mod. Phys. Lett.} A \textbf{5}, 2721 (1990).

\bibitem{Foot:1992ui}R. Foot {\it et al}, {\it J. Phys.} G \textbf{19}, 361 (1993).

\bibitem{Marinelli:1983nd}M. Marinelli and G. Morpurgo, \Journal{\PLB}{137}{439}{1984}.

\bibitem{Baumann:1988ue}J. Baumann {\it et al}, \Journal{\PRD}{37}{3107}{1988}.

\bibitem{Beda:2012zz}A. Beda {\it et al}, {\it Adv. High Energy Phys.} \textbf{2012}, 350150 (2012).

\bibitem{Raffelt}G. Raffelt, {\em Stars as laboratories for fundamental physics: The astrophysics of neutrinos, axions, and other weakly interacting particles} (University of Chicago Press, 1996).

\bibitem{Barbiellini:1987zz}G. Barbiellini and G. Cocconi, {\it Nature} \textbf{329}, 21 (1987).

\bibitem{Studenikin:2013my}
  A.~Studenikin,
  arXiv:1302.1168 [hep-ph].
\bibitem{Vogel:1989iv}P. Vogel and J. Engel, \Journal{\PRD}{39}{3378}{1989}.

\bibitem{Berestetskii:1979aa}V. Berestetskii {\it et al},
{\em Relativistic quantum theory} (Oxford, U.K.: Pergamon Press, 1979).

\bibitem{StudTern}A. Studenikin and A. Ternov, \Journal{\PLB}{608}{107}{2005}.

\bibitem{Stud2008}A. Studenikin, {\it J. Phys.} A \textbf{41}, 164047 (2008).

\bibitem{Balantsev:2010zw}
  I. Balantsev, Y. Popov and A. Studenikin,
  \Journal {\it J.\ Phys.\ A}{44}{255301}{2011}.

\bibitem{Balantsev:2012ep}
  I. Balantsev, A. Studenikin and I. Tokarev,
  {\it Phys. \ Part.\ Nucl.} {\bf 43}, 727  (2012).

\bibitem{Balantsev:2013aya}
  I. Balantsev, A. Studenikin and I. Tokarev,
  {\it Phys.\ Atom.\ Nucl.}  {\bf 76},  489 (2013).

 \bibitem{Studenikin:2012vi}
  A.~Studenikin and I.~Tokarev,
  arXiv:1209.3245 [hep-ph].

\bibitem{antares}M. Ageron {\it et al}, \Journal{\NIMA}{725}{60}{2013}.

\bibitem{glitch}P. Anderson and N. Itoh, {\it Nature} \textbf{256}, 25 (1975).

\bibitem{antiglitch}R. Archibald {\it et al}, {\it Nature} \textbf{497}, 591 (2013).

\bibitem{SN1987A}K. Hirata {\it et al}, {\it Phys.Rev.Lett.} \textbf{58} (1987) 1490.

\bibitem{pulsar_timing}R. Rosen {\it et al}, {\it Astrophys. J.} \textbf{768}, 85 (2013).

\end{thebibliography}
\end{document}